\documentclass[iop]{emulateapj}
\usepackage[para]{threeparttable}
\usepackage{graphics}

\usepackage{natbib}
\usepackage{float}

\begin{document}

\title{Analysis of the Early-Time Optical Spectra of SN 2011fe in M101}

\author{J.\ T.\ Parrent\altaffilmark{1,2}, 
D.\ A.\ Howell\altaffilmark{2,3},
B.\ Friesen\altaffilmark{4}, 
R.\ C.\ Thomas\altaffilmark{5}, 
R.\ A.\ Fesen\altaffilmark{1}, 
D.\ Milisavljevic\altaffilmark{6}, 
F.\ B.\ Bianco\altaffilmark{3,2}, 
B.\ Dilday\altaffilmark{2,3}, 
P.\ Nugent\altaffilmark{5,7},
E.\ Baron\altaffilmark{4,8,9},
I.\ Arcavi\altaffilmark{10}, S.\ Ben-Ami\altaffilmark{10}, D.\ Bersier\altaffilmark{11}, L.\ Bildsten\altaffilmark{3}, J.\ Bloom\altaffilmark{7}, Y.\ Cao\altaffilmark{12}, S.\ B.\ Cenko\altaffilmark{7}, A.\ V.\ Filippenko\altaffilmark{7}, A.\ Gal-Yam\altaffilmark{10}, M.\ M.\ Kasliwal\altaffilmark{13}, N.\ Konidaris\altaffilmark{12}, S.\ R.\ Kulkarni\altaffilmark{12}, N.\ M.\ Law\altaffilmark{14}, D.\ Levitan\altaffilmark{12}, K.\ Maguire\altaffilmark{15}, P.\ A.\ Mazzali\altaffilmark{16,17}, E.\ O.\ Ofek\altaffilmark{10}, Y.\ Pan\altaffilmark{15}, D.\ Polishook\altaffilmark{18},  D.\ Poznanski\altaffilmark{19}, R.\ M.\ Quimby\altaffilmark{20}, J.\ M.\ Silverman\altaffilmark{7}, A.\ Sternberg\altaffilmark{16}, M.\ Sullivan\altaffilmark{15}, E.\ S.\ Walker\altaffilmark{21}, D.\ Xu\altaffilmark{10}, C.\ Buton\altaffilmark{22}, and R.\ Pereira\altaffilmark{23}}

\altaffiltext{1}{6127 Wilder Lab, Department of Physics \& Astronomy, Dartmouth
                 College, Hanover, NH 03755, USA}
\altaffiltext{2}{Las Cumbres Observatory Global Telescope Network, Goleta, CA 93117, USA}
\altaffiltext{3}{Department of Physics, U.C. Santa Barbara, Santa Barbara, CA 93117, USA}
\altaffiltext{4}{Homer L. Dodge Department of Physics and Astronomy, University of Oklahoma, 440 W Brooks, Norman, OK 73019, USA}
\altaffiltext{5}{Physics Division, Lawrence Berkeley National Laboratory, 1 Cyclotron Road, Berkeley, CA 94720, USA}
\altaffiltext{6}{Harvard-Smithsonian Center for Astrophysics, 60 Garden St., Cambridge, MA 02138, USA}
\altaffiltext{7}{Department of Astronomy, University of California, Berkeley, CA 94720, USA}
\altaffiltext{8}{Hamburger Sternwarte, Gojenbergsweg 112, 21029 Hamburg, Germany}
\altaffiltext{9}{Physics Department, University of California, Berkeley, CA 94720 USA}
\altaffiltext{10}{Department of Particle Physics and Astrophysics, The Weizmann Institute of Science, Rehovot 76100, Israel}
\altaffiltext{11}{Astrophysics Research Institute, Liverpool John Moores University, Birkenhead CH41 1LD, UK}
\altaffiltext{12}{Cahill Center for Astronomy and Astrophysics, California Institute of Technology, 1200 E California Blvd., Pasadena, CA 91125, USA}
\altaffiltext{13}{Observatories of the Carnegie Institution for Science, 813 Santa Barbara St, Pasadena CA 91101 USA}
\altaffiltext{14}{Dunlap Institute for Astronomy and Astrophysics, University of Toronto, 50 St. George Street, Toronto M5S 3H4, Ontario, Canada}
\altaffiltext{15}{Department of Physics (Astrophysics), University of Oxford, Keble Road, Oxford OX1 3RH, UK}
\altaffiltext{16}{Max-Planck-Institut f\"ur Astrophysik, Karl-Schwarzschild-Strasse 1, D-85748 Garching, Germany}
\altaffiltext{17}{INAF-Osservatorio Astronomico, vicolo dell'Ossevatorio, 5, I-35122 Padova, Italy}
\altaffiltext{18}{Department of Earth, Atmospheric, and Planetary Sciences, Massachusetts Institute of Technology, Cambridge, MA 02139, USA}
\altaffiltext{19}{School of Physics and Astronomy, Tel-Aviv University, Tel-Aviv 69978, Israel}
\altaffiltext{20}{Kavli Institute for the Physics and Mathematics of the Universe, University of Tokyo, Kashiwa, Japan 277-8583}
\altaffiltext{21}{Scuola Normale Superiore, Piazza dei Cavlieri 7, 56126 Pisa, Italy}
\altaffiltext{22}{Physikalisches Institut Universit\"at Bonn (Bonn), Nussallee 12 53115 Bonn, Germany}
\altaffiltext{23}{Universit\`{e} de Lyon, F-69622 France; Universit\`{e} de Lyon 1; CNRS/IN2P3, Institut de Physique Nucl\`{e}aire de Lyon, France}

\begin{abstract}

  The nearby Type Ia supernova SN 2011fe in M101 ($cz=241$ km s$^{-1}$) provides a unique opportunity to study the early evolution of a ``normal'' Type Ia supernova, its compositional structure, and its elusive progenitor system. We present 18 high signal-to-noise spectra of SN
  2011fe during its first month beginning 1.2 days
  post-explosion and with an average cadence of 1.8 days. This gives a
  clear picture of how various line-forming species are distributed
  within the outer layers of the ejecta, including that of unburned material (C+O).  We follow the evolution of
  \ion{C}{2} absorption features until they diminish near maximum
  light, showing overlapping regions of burned and unburned material
  between ejection velocities of 10,000 and 16,000 km s$^{-1}$. This
  supports the notion that incomplete burning, in addition to progenitor scenarios,
  is a relevant source of spectroscopic diversity among SNe Ia. The
  observed evolution of the highly Doppler-shifted \ion{O}{1}
  $\lambda7774$ absorption features detected within five days
  post-explosion indicate the presence of \ion{O}{1} with expansion velocities from 11,500 to
  21,000 km s$^{-1}$. The fact that some \ion{O}{1} is present above
  \ion{C}{2} suggests that SN 2011fe may have had an appreciable amount of unburned oxygen
  within the outer layers of the ejecta.

\end{abstract}


\section{Introduction}

SN 2011fe in M101 is the closest (6.9 Mpc) Type Ia supernova (SN Ia)
in 25 years. It was discovered only 11 hours after explosion, the
earliest of any SN Ia \citep{Nugent11} and has revealed more about its progenitor than any
other recent SN Ia. In the canonical picture of a SN Ia, a CO white dwarf (WD)
gains mass until it approaches the Chandrasekhar mass, either by mass
transfer from a nondegenerate companion, or by merging
with another WD (for a review, see \citealt{Howell11}). 

The merger of two WDs has been strongly suspected in the case
of ``super-Chandra'' SNe~Ia, those that are so luminous they
appear to require a progenitor above the Chandrasekhar mass
\citep{Howell06,Scalzo10}.  Unlike most SN~Ia, these subtypes are known for showing conspicuous and relatively long-lasting spectroscopic signatures of carbon in their outer layers, possibly left over from the violent CO-WD
merger \citep{Silverman11}.  But this scenario cannot explain all SNe~Ia as some show
narrow lines of sodium from circumstellar material (CSM), thought to be
from a nondegenerate secondary star in a single-degenerate system
(\citealt{Dilday12} and references therein).
\citet{Sternberg11} estimate that at least 20\% of SNe~Ia in spiral
galaxies arise from this channel.  Furthermore,  it remains to be seen if circumstellar sodium is concurrent with carbon detections.  Could abundance studies of carbon lines reveal the progenitor system?

With SN 2011fe, it was confirmed observationally for the first time that the exploding star was a WD
\citep{Nugent11,Bloom12}.  Pre-explosion \emph{Hubble Space Telescope}
images revealed that any secondary star could not be a luminous red
giant \citep{WLi11}.  Furthermore, the lack of an observed shock,
expected when the ejecta hits any nondegenerate secondary
\citep{Kasen10}, suggests that the secondary could not be a red
giant \citep{Nugent11}, and probably not any nondegenerate star
\citep{Bloom12}.  Nondetections of radio and X-ray emission place
strong limits on the amount of CSM,
disfavoring most single-degenerate hypotheses, but consistent with the
merger of two WD stars \citep{Horesh12, Chomiuk12, Margutti12}. 

In addition, thorough and complete spectroscopic observations of SN 2011fe offer a chance to study the distribution of ejected material. For instance, as many as a third of SNe Ia show evidence of unburned material if spectra are obtained early
enough \citep{Parrent11,Thomas11,Folatelli11,SF12,Blondin12}.  Carbon features may not always, if at all, be the signature of a
merger \citep{Scalzo10}, but rather signal some difference in the explosion physics of the primary WD \citep{Baron03,Thomas07}.  Abundance studies of unburned material can be coupled with constraints on material that is synthesized in the explosion, as some explosion models predict the complete burning of the WD, leaving no unburned carbon \citep{Kasen09,Maeda10}.     

To investigate these questions it is necessary to combine our limited
progenitor knowledge with a map of the elemental distribution of ejected material in SNe,
including unburned material. Here we present 18 high signal-to-noise (S/N) optical
spectra of SN 2011fe during its first month, starting one day after
explosion.  We fit all spectra with the automated
spectrum synthesis code, \texttt{SYNAPPS}\footnote[24]{Derived from
  \texttt{SYNOW}, see \citet{Branch05} and \citet{Thomas11a}.}, to
trace the compositional structure of the ejected material. 

\section{Observations}

\begin{center}
\begin{figure*}[ht,scale=1.5]
\includegraphics*[scale=0.8]{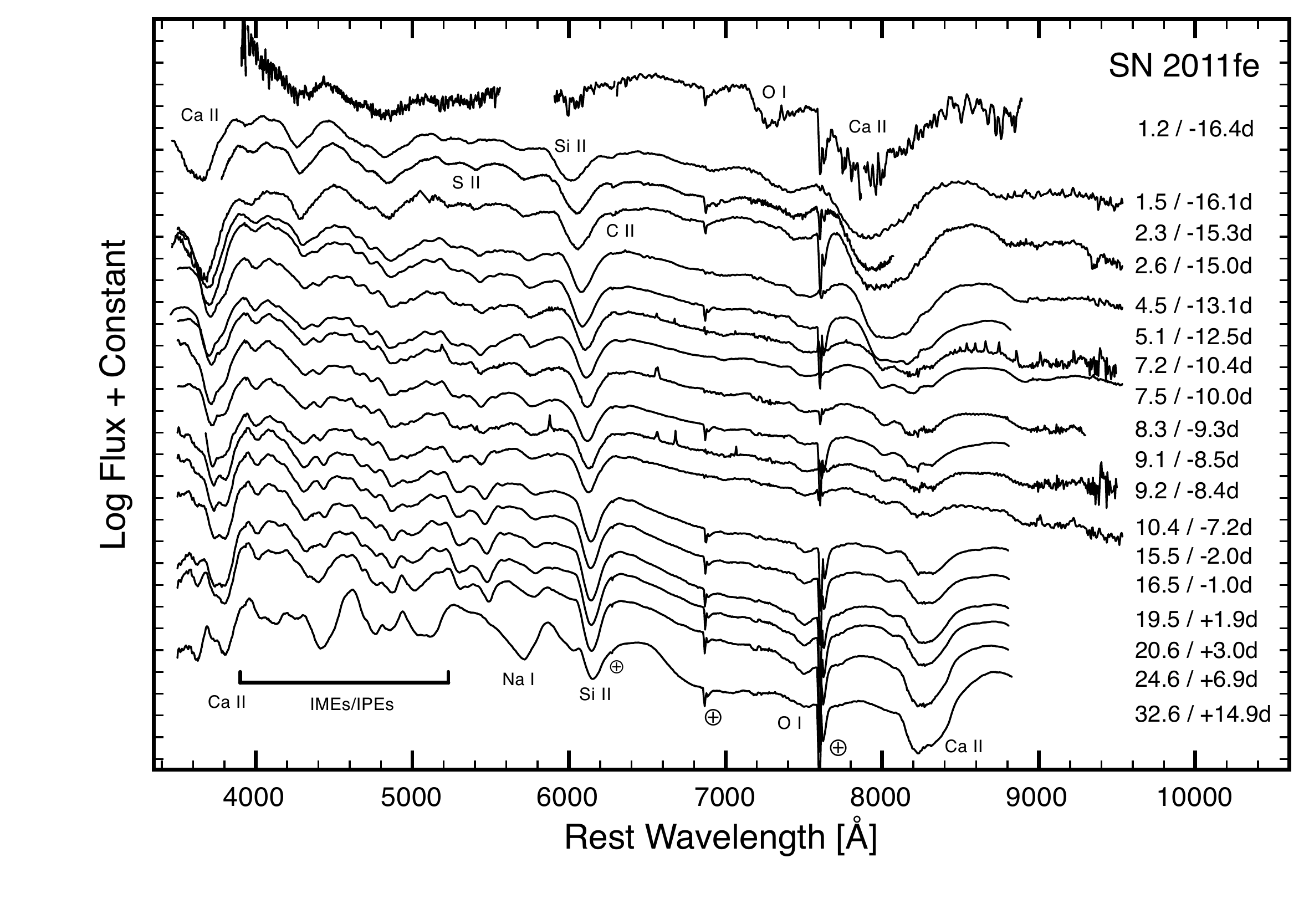}
\centering
\caption{Optical spectra of SN 2011fe during the first month of observations, labeled with respect to an explosion date of 2011 August UT 23.69 and a B-band maximum light date of 2011 September UT 10.3. Most of the data span 3500 $-$ 9700 \AA, roughly every 2 days between the pre-maximum and post-maximum phases. Line identifications are labeled for some of the most conspicuous features. The $\earth$- symbol indicates the location of telluric features near 6279, 6880, and 7620 \AA.}
\label{fig:data} 
\end{figure*}
\end{center}

On 2011 August 24.167 (UT dates are used throughout this Letter), the Palomar Transient Factory (PTF) discovered SN
2011fe in M101. The SN was discovered early enough
that the time of explosion, $t_{\rm exp}$, could be estimated to within
$\sim 20$ minutes of 2011 August 23.69 \citep{Nugent11}.  
The SN reached a $B$-band peak
brightness on September 10.3 (\citealt{Bianco12}). Optical spectra were obtained on eight
different telescopes (Table 1, Figure~\ref{fig:data}). Reductions were done with IRAF\footnote[25]{IRAF is distributed by the National
  Optical Astronomy Observatory, which is operated by the Association
  of Universities for Research in Astronomy, Inc., under cooperative
  agreement with the National Science Foundation (NSF).}.

\section{Spectroscopic Analysis}

From the onset, the spectroscopic features of SN 2011fe appear broad and comprised of P-Cygni absorption/emission profiles of intermediate-mass elements (IMEs) and iron-peak elements (IPEs) typically seen in supernova spectra \citep{BP73,Flipper97}. Over time the features narrow, depicting an evolution similar to ``normal'' subtype objects. Most notably the spectra exhibit: (1) high velocity features (HVFs; $v >$ 16,000 km s$^{-1}$) of \ion{Si}{2}, \ion{Ca}{2} and \ion{Fe}{2} absorption lines from the onset of the explosion, (2) unburned material in the form of a \ion{C}{2} $\lambda$6578 absorption feature, and (3) \ion{O}{1} $\lambda$7774 at velocities higher than previously detected for a SN Ia \citep{Nugent11}. From the lack of conspicuous \ion{Ti}{2}/\ion{Fe}{3} absorption features sometimes observed in sub- and over-luminous SNe Ia, respectively, as well as the slowly decreasing blueshift of the \ion{Si}{2} 6150 \AA\ feature, our spectroscopic observations of SN 2011fe in Figure~\ref{fig:data} are consistent with characteristics of a ``low velocity gradient''/``core normal'' subtype (see \citealt{Blondin12} and references therein).

\subsection{Line Identifications: \texttt{SYNAPPS}} 

Given that supernova expansion velocities are $\sim$10$^{4}$ km
s$^{-1}$, Doppler line blending can be an obstacle
for mapping the underlying compositional structure. \texttt{SYNAPPS}
is a useful tool for identifying line features in photospheric phase
spectra through its empirical spectrum synthesis procedure
which requires no \emph{a priori} compositional structure as input.
Under the assumption of pure-resonance line transfer and a sharp
blackbody-emitting photosphere, an atomic line list is used to
construct a basic spectrum from a pseudo-blackbody continuum level.
The optical depth profile for a line is modeled as $\tau(v) = \tau_{ref} \times \exp(\frac{v_{ref} - v}{v_{e}})$, where $v$ is the expansion velocity, $v_{ref}$ is an arbitrary normalization velocity, $v_{e}$ is the e-folding velocity, and $\tau_{ref} = \tau(v_{ref})$
\citep{Thomas11a}.

Because the sources of line formation influence nearly all regions of the spectrum
from a wide range of depths within the ejecta, we model the spectrum
approximately by treating the minimum reference velocity, $v_{\rm min}$, for every ion as a free
parameter. For a given set of ions in a fit, we set the photospheric velocity,
$v_{\rm phot}$, equal to the lowest measured $v_{\rm min}$. This follows the notion
that not all blended absorption minima coincide with the
Doppler-shifted location of the photosphere over time \citep{Zeb08}. Furthermore, we use a quadratic
warping function to simplify the treatment of the continuum, calibration errors, and uncorrected reddening. Consequently, our fits are constrained by how we are able to best match absorption
minima and the relative strengths of all features.   

\begin{center}
\begin{figure*}[ht,scale=1.5]
\centering
\includegraphics*[scale=0.6]{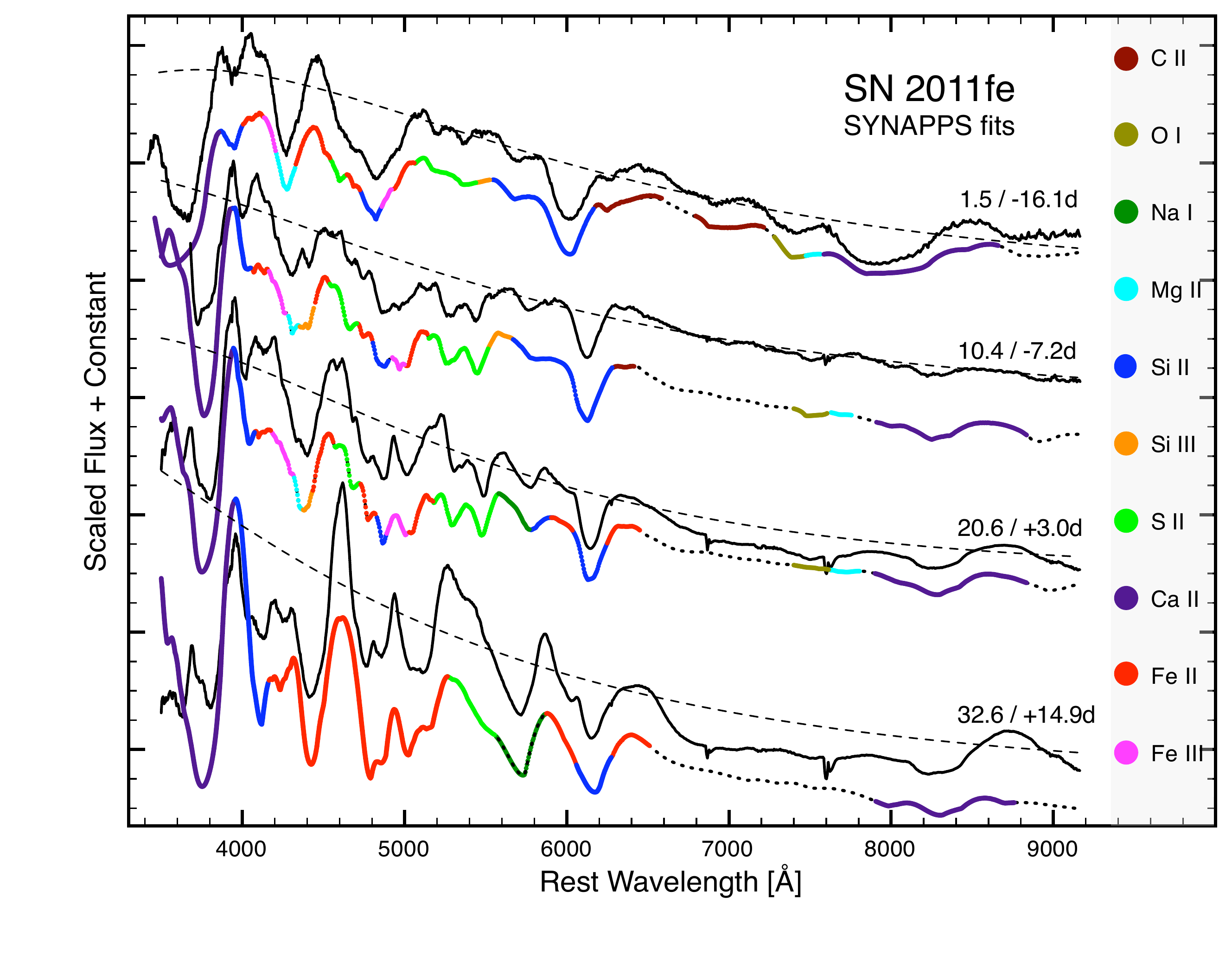}
\caption{Synthetic spectra comparisons are shown for days 1.5, 10.4, 20.6 and 32.6 post-explosion ($-$16.1, $-$7.2, $+$3.0 and $+$14.9 days with respect to maximum light). The dashed line represents the fitted continuum level. We color various pieces of our fits to indicate the dominant ion we attribute to the features. The black dotted line of the fit indicates where we do not attribute features to any of the assumed ions.}
\label{fig:fits}
\end{figure*}
\end{center}

For the spectra in Figure~\ref{fig:data}, we converged a series of \texttt{SYNAPPS} fits. In Figure~\ref{fig:fits} we plot the spectra for the days 1.5, 10.4, 20.6 and 32.6 post-explosion, as they best document the relevant temporal changes on which we focus our attention. For each epoch, we compare our \texttt{SYNAPPS} fit and over-plot the continuum fit for reference. We have loosely highlighted various regions of the fits to indicate which parts of the data we use to infer, and thereby constrain, the presence of particular ions. For example, the two broad absorption features centered around 4300 and 4800 \AA\ are complex blends of three or more ions. When we remove any one of these ions from the fit, we are unable to compensate for it with any reasonable perturbations of the remaining parameters. That is to say, while one or two ions may dominate it, the shape of a particular blended feature is still subject to the minor but observable influences of other species.

For the 1.5 day fit we set $v_{phot}$ = 16,000 km s$^{-1}$ and include photospheric components of \ion{C}{2}, \ion{O}{1}, \ion{Mg}{2}, \ion{Si}{2}, \ion{S}{2}, \ion{Ca}{2}, and \ion{Fe}{3}. In addition, we include \emph{higher} velocity components of \ion{O}{1}, \ion{Si}{2}, \ion{Ca}{2}, and \ion{Fe}{2} \emph{above} 16,000 km s$^{-1}$ in order to follow the evolution of the corresponding HVFs detected during the earliest observed epochs \citep{Nugent11}. Overall, many of the features are accounted for in our subsequent fits, and the relative fluxes and respective minima match nicely. 

Over the course of their detection, we model the \ion{Ca}{2} HVFs as material ``detached'' 10,000 km s$^{-1}$ above the photospheric line forming regions. It is clear when the \ion{Ca}{2} HVFs disappear for SN 2011fe at $\sim$5 days after $t_{expl}$; however, the same is not true for \ion{Si}{2}. We model \ion{Si}{2} as present in both photospheric and higher velocity regions, despite the separation of 2000 km s$^{-1}$ near maximum light. It is sometimes clear that the 6150 \AA\ feature requires two distinctly separate components of \ion{Si}{2} for a best fit (most notably SN 2009ig; see \citealt{Foley12}). Recently however, \citet{Smith11} presented spectropolarimetry of SN 2011fe showing two \ion{Si}{2} absorption line polarizations at $\sim$7300 and 12,000 km s$^{-1}$ just after maximum light (see their Figure 3). During this phase our two components of \ion{Si}{2} are at 8000 and 11,000 km s$^{-1}$. While the overlap is not exact, the finding of a two-component \ion{Si}{2} feature is consistent with the spectropolarimetry.

The set of ions mentioned above is primarily the same used until maximum light, at which point \ion{Na}{1}, \ion{S}{2}, and \ion{Fe}{2} begin to account best for most of the features. For SN 2011fe, while using only \texttt{SYNAPPS}, we cannot definitively infer the presence of ions such as cobalt and nickel from the first month of optical observations alone, i.e., species responsible for powering SN Ia lightcurves \citep{Colgate69}. However, early-phase broadband photometry, as well as spectroscopic analyses of infrared data, could help in this regard \citep{Marion09,Piro12}.

\subsection{Spectroscopic Modeling: \texttt{SYNAPPS}}

The result of any series of \texttt{SYNAPPS} fits is the estimation of velocities spanned by a particular set of ions. Therefore, we can use spectra in two different ways, as seen in Figure~\ref{fig:onion}. From our time-series fits, in Figure~\ref{fig:onion}(a) we show our $v_{min}(t_{obs})$ estimates for each of the proposed ions in \S3.1. A clear separation between outer and inner regions of ejected material can be seen. The material near photospheric velocities that span roughly $7000 - 16,000$ km s$^{-1}$ includes \ion{Na}{1}, \ion{C}{2}, \ion{O}{1}, \ion{Mg}{2}, \ion{Si}{2}, \ion{S}{2}, \ion{Ca}{2}, and \ion{Fe}{3}, whereas most of the higher velocity material between 16,000 and 30,000 km s$^{-1}$ is comprised of \ion{O}{1}, \ion{Ca}{2}, \ion{Si}{2}, and \ion{Fe}{2}.  

Since we treat $v_{min}$ as a free parameter, there is some dispersion among ions near the modeled photosphere for a given epoch (black dashed line in Figure~\ref{fig:onion}(a)). However, \texttt{SYNAPPS} fits are generally good to within 250 km s$^{-1}$ and the scatter of our inferred $v_{min}$ is no less than 500 km s$^{-1}$. Therefore we consider the dispersion consistent with what should be expected in an environment of multiple line forming regions. 

When we take into account the Doppler-widths of our fits, we are able to place tighter constraints on the location of species by constructing a velocity-scaled map that traces how far each ion extends. In Figure~\ref{fig:onion}(b) we plot a single normalized absorption profile for each ion, summed over the first month. Clearly, a photospheric and higher velocity region can be seen. In addition, Figure~\ref{fig:onion}(b) yields some insight into where species are likely to be present, with respect to one another, throughout the ejecta. This is particularly important for investigating unburned material since its location with respect to burned material can be used to constrain the details of the explosion nucleosynthesis. 

Figure~\ref{fig:carbon} highlights regions of interest for tracing
unburned material with \ion{C}{2} $\lambda$6578 and potentially
\ion{O}{1} $\lambda$7774 (see also Figure~\ref{fig:onion}). For our
\ion{C}{2} fits, we are able to follow the suspected absorption
feature as it evolves across the emission component of the \ion{Si}{2}
$\lambda$6355 line over time. This implies that unburned carbon is
present at least between 10,000 and 16,000 km s$^{-1}$. Our \ion{O}{1} fits in
Figure~\ref{fig:carbon} require both photospheric and higher velocity
components of \ion{O}{1} for the 1.2, 1.5, 2.6 and 4.5 day spectra as
the \ion{O}{1} HVF disappears quickly and within 5 days after
$t_{exp}$. Within the time-frame of our observations, \ion{C}{2}
$\lambda$6578 is present from 1.5 to 20.6 days after $t_{exp}$, i.e.,
from the onset and until at least maximum light. The same is true for
\ion{O}{1}; however we find that \ion{O}{1} initially extends at least
5000 km s$^{-1}$ \emph{above} \ion{C}{2}, within the outermost layers. This suggests the \ion{O}{1}
detected during the earliest epochs is likely to be unburned
progenitor material rather than a product of carbon burning in the SN.

\section{Discussion \& Conclusions}

\begin{figure*}[tbp]
\centering
\includegraphics*[scale=1.0]{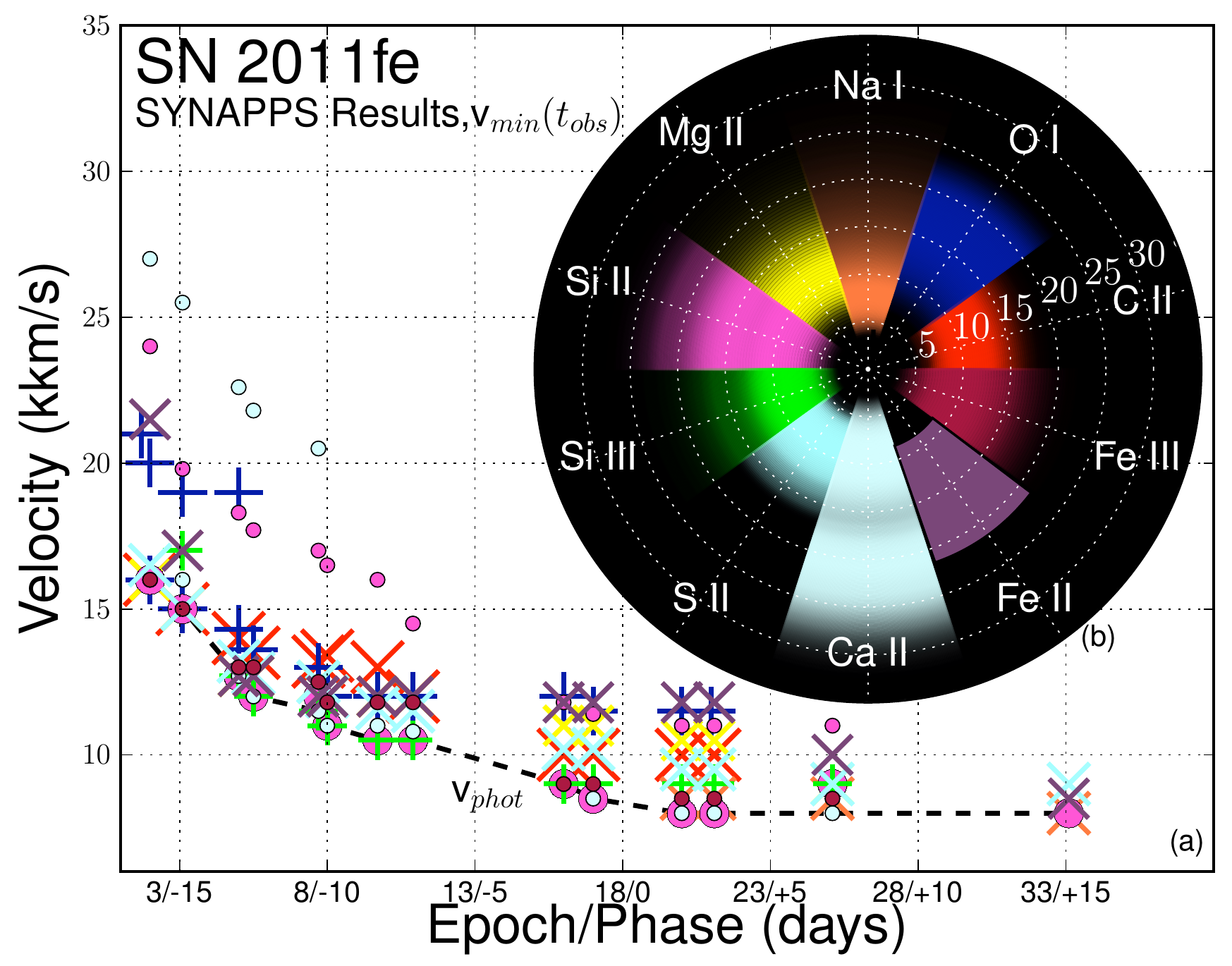}
\caption{Symbols are for clarity within the plot and different ions are represented by different colors, as labeled in panel (b). (a) Based on our time-series of synthetic \texttt{SYNAPPS} spectra, $v_{min}$ estimates of each ion are plotted versus time. The dashed line represents the location of our best modeled photosphere. (b) Here we reproduce panel (a), but include the Doppler widths of our fits. Color intensities are mapped to the normalized sum of a single absorption line component for each ion in every fit. Therefore, panel (b) only depicts where light has been \emph{scattered} by a particular ion. Due to telluric features, we have only included our fitted Doppler widths for the high velocity \ion{O}{1} component. The numbered rings (dashed-white) are in increments of 5 kkm s$^{-1}$.}
\label{fig:onion}
\end{figure*}

\begin{figure*}
\centering
\includegraphics*[scale=0.8]{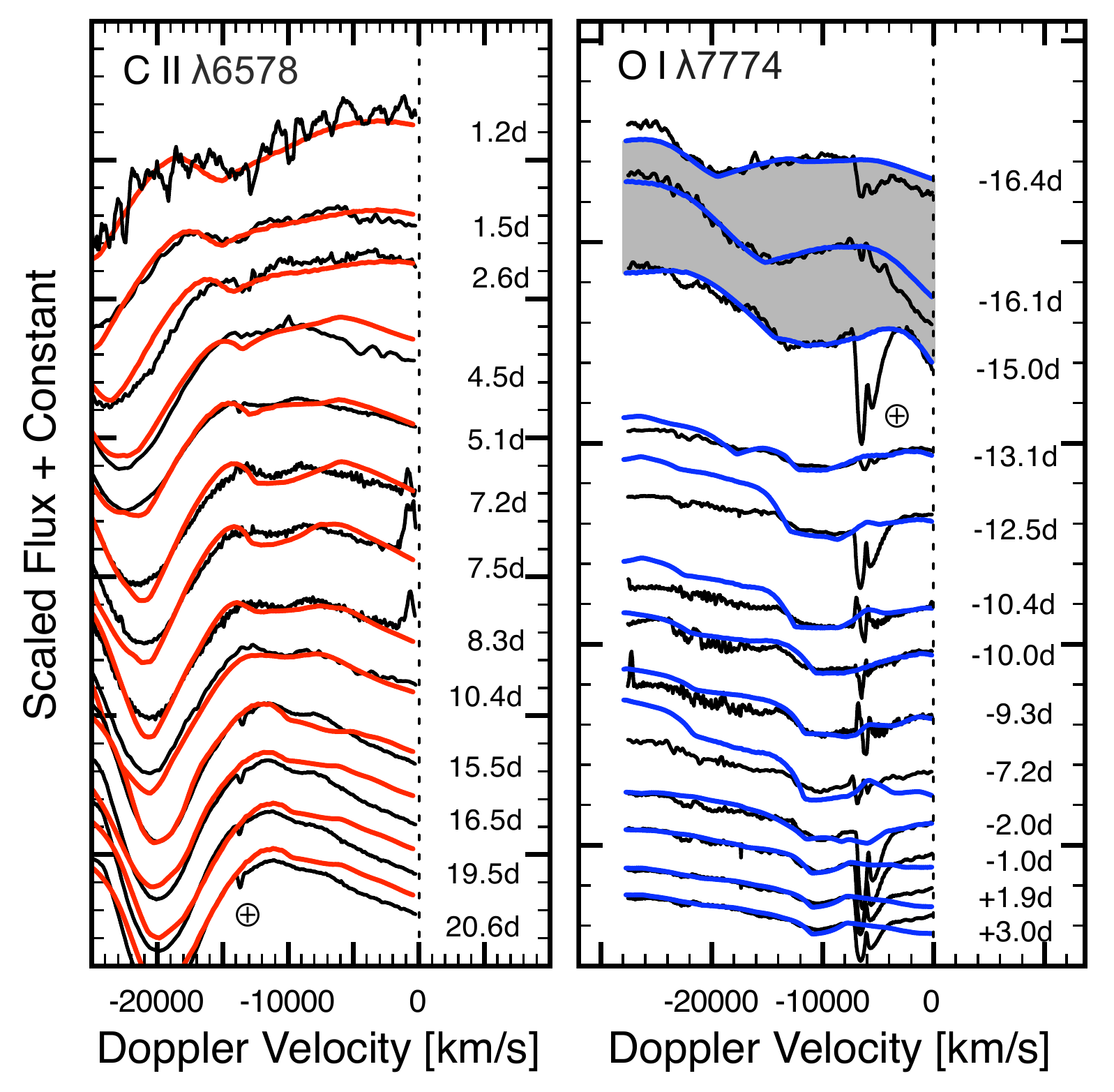}
\caption{``Snap-shots'' of unburned material, i.e., spectroscopic signatures of unburned material as best traced by \ion{C}{2} $\lambda$6578 and possibly \ion{O}{1} $\lambda7774$. On the left, we compare our \texttt{SYNAPPS} fits (red line) to the observations (black line) that show a feature near 6300 \AA. At 1.2 days after $t_{expl}$ the \ion{C}{2} absorption minimum corresponds to a Doppler velocity of 16,000 km s$^{-1}$, at the upper reaches of photospheric velocities. On the right, we plot the same but instead for the \ion{O}{1} $\lambda$7774 absorption feature. Our fits (blue line) place \ion{O}{1} in both photospheric and higher velocity regions. The gray band highlights the early evolution of the \ion{O}{1} feature and the $\earth$- symbol indicates the location of telluric features.}
\label{fig:carbon}
\end{figure*}

Thanks to the proximity of SN 2011fe, we have been able to obtain frequent high S/N spectra and use \texttt{SYNAPPS} to produce what may be the best-ever map of the material distribution in a ``normal'' SN Ia for the first month of its evolution. For SN 2011fe, it is important to note that, apart from differences in relative expansion velocities (Figure~\ref{fig:onion}(a)), the spectra do not vary significantly from one epoch to another. Furthermore, as in most SNe Ia, in SN 2011fe we find \ion{O}{1}, \ion{Mg}{2}, \ion{Si}{2},
\ion{S}{2}, \ion{Ca}{2}, and \ion{Fe}{2} present
from 16,000 km s$^{-1}$ or more down to about 7000 km s$^{-1}$ (later
spectra will probe deeper). This implies similar compositional constituents over a large region of ejected material and is supported by our \texttt{SYNAPPS} fitting results (Figure~\ref{fig:onion}(b)). 

Stretching to nearly 30,000 km s$^{-1}$, high velocity regions of silicon and calcium are inferred from pre-maximum light spectra. A region of sodium is inferred to be present at high velocities as well, but is detected post-maximum light when the outer layers begin to cool. However, rather than large-scale asymmetries, as predicted by \citet{MaedaNature}, these high velocity features may trace only a small amount of material, as little as $10^{-7} M_\odot$ in the case of \ion{Ca}{2} \citep{Hatano99,Mazzali05,Tanaka08}. Such a small inferred Ca mass supports the notion that strong features in SN spectra do \emph{not} always correspond to large masses of line-forming ejecta.

We also find \ion{Si}{3} and \ion{Fe}{3} present at early times (before 3 days
after maximum light), indicating that some parts of the ejecta achieved high temperatures, although the entire photosphere was not excessively hot.
Even a day after explosion the spectra were not particularly blue, nor
were they dominated by high-ionization species.  Various scenarios
that predict an early shock from interaction with a
nondegenerate secondary star \citep{Kasen10}, a torus of material from
a disrupted secondary \citep{Scalzo10}, or CSM, are
not supported by the data.  This is consistent with a lack of radio
and X-ray detections \citep{Horesh12, Chomiuk12, Margutti12}, although
the spectra here start a day earlier, providing additional constraints. 

Regarding unburned material, other SNe Ia show distributions of unburned carbon spanning a range of velocities. This suggests that carbon in some SNe Ia arises from incomplete burning, in contrast to the predictions of some delayed detonation models and merger scenarios \citep{Marion09,Fryer10,Roepke12}. In general, due to the rarity of early SN Ia observations, little is certain about the conditions under which
carbon is observable in all subtypes.  Noise makes weak carbon lines
difficult to detect, especially when supernovae are faint at early
times, and different velocities cause it to be blended with other
features. Furthermore, SN Ia subtypes have
different temperatures \citep{Nugent95}, and these change with time,
affecting whether carbon signatures exists as \ion{C}{1}, \ion{C}{2}, or
\ion{C}{3}.  Thus carbon may be fairly common in SNe Ia, but only
apparent if spectra are obtained early \citep{Parrent11,Thomas11,Folatelli11,SF12,Blondin12}. In addition, asymmetry may play a part in whether carbon is detected given that both explosive burning and WD mergers are inherently asymmetric processes \citep{Livio11, Pakmor12}. Spectropolarimetry and late-time observations, which are only possible for bright, nearby SNe like SN 2011fe, can shed light on these links.

For the carbon in SN 2011fe, we find \ion{C}{2} to be strongest in the earliest spectra,
suggesting it is mostly concentrated in the outer layers. The \ion{C}{2} $\lambda$6578 detection persists until just before maximum light, although at later times it is very weak and would have been missed in any but the
highest S/N spectra. Note, however, its presence {\it throughout} regions of freshly synthesized material shows that it is indeed distributed within the ejecta and is not just concentrated within the outermost layers. Because the high
velocity \ion{O}{1} is located above the outermost reaches of the \ion{C}{2}, it is likely unburned progenitor material \citep{Nugent11}. 

For the purposes of investigating spectroscopic diversity, it is convenient that SN 2011fe exhibits the characteristic signatures of a ``normal'' SN Ia. Future comparative studies of SNe Ia will benefit from these results and that of other well observed objects. The map of elements in SN 2011fe
can now be used as input for model atmosphere codes, such as
PHOENIX (see \citealt{Hauschildt99} and references
therein), enabling us to better determine abundances, test theories of
the explosion physics, and perhaps gain further insight into
progenitor scenarios. 

\acknowledgements

This work was supported by the Las Cumbres Observatory Global Telescope Network. This research used resources of the National Energy Research Scientific Computing Center, which is supported by the Office of Science of the U.S. Department of Energy under contract no. DE-AC02-05CH11231.

Observations were obtained as part of the Palomar Transient Factory project. The TNG is operated by the Fundacion Galileo Galilei, the LT and WHT are operated by the Isaac Newton Group, and are located in the Spanish Observatorio of the Roque de Los Muchachos of the Instituto de Astrofisica de Canarias. Some observations were obtained at the Gemini Observatory (GN-2011B-Q-215), which is operated by the Association of Universities for Research in Astronomy, Inc., under a cooperative agreement with the NSF on behalf of the Gemini partnership. Other observations were obtained with the 200 inch telescope at Palomar Observatory, the UH 2.2-m telescope, which is part of the Nearby Supernova Factory II project, and the Lick 3-m telescope, which is operated by the University of California. Some of the data presented herein were obtained at the W. M. Keck Observatory.

\bibliographystyle{apj}

\clearpage

\renewcommand{\thefootnote}{\alph{footnote}}

\begin{table}
\begin{threeparttable}[b]
\begin{tabular}{ccccccl}
 \tableline\tableline
 UT & MJD  & Epoch\tablenotemark{a} & Phase\tablenotemark{b} & Telescope & Exp. Time & Range \\
 Date & 55,000 & days & days & + Instrument & (s) & (\AA) \\
 \tableline
2011 Aug 24 & 797.86 & 1.2   & $-$16.4 & LT+FRODOSpec  & 1800 & 3900 $-$ 9000\tablenotemark{$\dagger$} \\ 
2011 Aug 25 & 798.16 & 1.5   & $-$16.1 & Lick3-m+Kast  & 1200 & 3420 $-$ 10290\tablenotemark{$\dagger$} \\        %\nodata &  
2011 Aug 25 & 798.95 & 2.3   & $-$15.3 & TNG+DOLORES & 300 & 3780 $-$ 8070 \\         %\nodata & 
2011 Aug 26 & 799.26 & 2.6   & $-$15.0 & UH88+SNIFS & 300 & 3300 $-$ 9700 \\            %\nodata & 
2011 Aug 28 & 801.19 & 4.5   & $-$13.1 & P200+DBSP & 120 & 3300 $-$ 10000 \\        %\nodata & 
2011 Aug 29 & 801.74 & 5.1   & $-$12.5 & GN+GMOS  & 120($\times$4) & 3500 $-$ 9700 \\       %6 & 
2011 Aug 30 & 803.86 & 7.2   & $-$10.4 & WHT+ISIS  & 60($\times$2) & 3500 $-$ 9500 \\       %\nodata & 
2011 Aug 31 & 804.23 & 7.5   & $-$10.0 & Keck1+LRIS  & 45 & 3050 $-$ 10200 \\       %\nodata & 
2011 Aug 31 & 804.94 & 8.3   & $-$9.3 & WHT+ISIS  & 60($\times$2) & 3500 $-$ 9300 \\       %\nodata &
2011 Sep 1 & 805.75 & 9.1   & $-$8.5 & GN+GMOS & 60($\times$2) & 3500 $-$ 9700 \\       %6 & 
2011 Sep 1 & 805.90 & 9.2   & $-$8.4 & WHT+ISIS  & 120 & 3500 $-$ 9300 \\      %\nodata & 
2011 Sep 3 & 807.13 & 10.4 & $-$7.2 & P200+DBSP & 120 & 3680 $-$ 9700 \\       %\nodata & 
2011 Sep 8 & 812.23 & 15.5 & $-$2.0 & GN+GMOS &  60($\times$2) & 3500 $-$ 9700 \\      %6 
2011 Sep 9 & 813.23 & 16.5 & $-$1.0 & GN+GMOS & 60($\times$2) & 3500 $-$ 9700 \\      %6 & 
2011 Sep 12 & 816.22 & 19.5 & +1.9 & GN+GMOS &  60($\times$2) & 3500 $-$ 9700 \\       %6 
2011 Sep 13 & 817.27 & 20.6 & +3.0 & GN+GMOS & 60($\times$2) & 3500 $-$ 9700 \\      %6 & 
2011 Sep 17 & 821.27 & 24.6 & +6.9 & GN+GMOS & 60($\times$2) & 3500 $-$ 9700 \\      %6 & 
2011 Sep 25 & 829.27 & 32.6 & +14.9 & GN+GMOS & 60($\times$2) & 3500 $-$ 9700 \\     %6 & 
 \tableline
\end{tabular}
\begin{tablenotes}
\item [a] Relative to an estimated explosion JD of 2,455,797.187 $\pm$ 0.014 \citep{Nugent11}. \\
\item [b] Relative to an estimated M$_{b, max}$ JD of 2,455,814.28 $\pm$ 0.02 (\citealt{Bianco12}). \\
\item [$\dagger$] Spectra presented in \citealt{Nugent11}.
\end{tablenotes}
\end{threeparttable}
\end{table}

\clearpage
\newpage

\end{document}